\documentclass[12pt,preprint2]{emulateapj}

\usepackage{lineno}
\usepackage{natbib}
\usepackage{amsmath}
\usepackage{graphicx}
\usepackage{wasysym}
\usepackage{txfonts}
\usepackage{xspace}
\usepackage{url}
\usepackage{textcomp}
\usepackage{multirow}
\usepackage{lipsum}

\newcommand{\vela}{Vela~X-1\xspace}

\newcommand{\cep}{Cep~X-4\xspace}
\newcommand{\gx}{GX~301$-$2\xspace}
\newcommand{\vo}{V0332+53\xspace}
\newcommand{\her}{Her~X-1\xspace}

\newcommand{\swift}{\textsl{Swift}\xspace}

\newcommand{\suz}{\textsl{Suzaku}\xspace}

\newcommand{\xte}{\textsl{RXTE}\xspace}
\newcommand{\nustar}{\textsl{NuSTAR}\xspace}

\newcommand{\snr}{S/N\xspace}

\newcommand{\msun}{\ensuremath{\text{M}_{\odot}}\xspace}

\newcommand{\feka}{\ensuremath{\mathrm{Fe}~\mathrm{K}\alpha}\xspace}

\newcommand{\nh}{\ensuremath{{N}_\mathrm{H}}\xspace}

\shorttitle{Distorted cyclotron line profile in \cep}
\shortauthors{F\"urst et al.}

\begin{document}

\title{Distorted cyclotron line profile in Cep~X-4 as observed by \nustar}

\author{F.~F\"urst\altaffilmark{1}}
\author{K.~Pottschmidt\altaffilmark{2,3}}
\author{H.~Miyasaka\altaffilmark{1}}
\author{V.~Bhalerao\altaffilmark{4}}

\author{M.~Bachetti\altaffilmark{5}}
\author{S.~E.~Boggs\altaffilmark{6}}
\author{F.~E.~Christensen\altaffilmark{7}}
\author{W.~W.~Craig\altaffilmark{6,8}}
\author{V.~Grinberg\altaffilmark{9}}
\author{C.~J.~Hailey\altaffilmark{10}}
\author{F.~A.~Harrison\altaffilmark{1}}
\author{J.~A.~Kennea\altaffilmark{11}}
\author{F.~Rahoui\altaffilmark{12,13}}
\author{D.~Stern\altaffilmark{14}}
\author{S.~P.~Tendulkar\altaffilmark{1}}
\author{J.~A.~Tomsick\altaffilmark{6}}
\author{D.~J.~Walton\altaffilmark{14,1}}
\author{J.~Wilms\altaffilmark{15}}
\author{W.~W.~Zhang\altaffilmark{3}}

\altaffiltext{1}{Cahill Center for Astronomy and Astrophysics, California Institute of Technology, Pasadena, CA 91125, USA}
\altaffiltext{2}{CRESST, Department of Physics, and Center for Space Science and Technology, UMBC, Baltimore, MD 21250, USA}
\altaffiltext{3}{NASA Goddard Space Flight Center, Greenbelt, MD 20771, USA}
\altaffiltext{4}{Inter-University Center for Astronomy and Astrophysics,  Ganeshkhind, Pune 411007, India}
\altaffiltext{5}{Osservatorio Astronomico di Cagliari, 09047 Selargius (CA), Italy}
\altaffiltext{6}{Space Sciences Laboratory, University of California, Berkeley, CA 94720, USA}
\altaffiltext{7}{DTU Space, National Space Institute, Technical University of Denmark, 2800 Lyngby, Denmark} 
\altaffiltext{8}{Lawrence Livermore National Laboratory, Livermore, CA 94550, USA}
\altaffiltext{9}{Massachusetts Institute of Technology, Kavli Institute for Astrophysics, Cambridge, MA 02139, USA}
\altaffiltext{10}{Columbia Astrophysics Laboratory, Columbia University, New York, NY 10027, USA}
\altaffiltext{11}{Department of Astronomy \& Astrophysics, The Pennsylvania State University, University Park, PA 16802, USA}
\altaffiltext{12}{European Southern Observatory,  85748 Garching bei M\"{u}nchen, Germany} 
\altaffiltext{13}{Department of Astronomy, Harvard University, Cambridge, MA 02138, USA} 
\altaffiltext{14}{Jet Propulsion Laboratory, California Institute of Technology, Pasadena, CA 91109, USA}
\altaffiltext{15}{Dr. Karl-Remeis-Sternwarte and ECAP,  University of Erlangen-Nuremberg, 96049 Bamberg, Germany} 


%

\begin{abstract}
We present spectral analysis of \nustar and \swift observations of \cep during its outburst in 2014. We observed the source once during the peak of the outburst and once during the decay, finding good agreement in the spectral shape between the observations. We describe the continuum using a powerlaw with a Fermi-Dirac cutoff at high energies. \cep has a very strong cyclotron resonant scattering feature (CRSF) around 30\,keV.
 A simple absorption-like line with a Gaussian optical depth or a pseudo-Lorentzian profile both fail to describe the shape of the CRSF accurately, leaving significant deviations at the red side of the line. We characterize this asymmetry with a second absorption feature around 19\,keV.  The line energy of the CRSF, which is not influenced by the addition of  this feature, shows a small but significant positive luminosity dependence. With luminosities between (1--6)$\times10^{36}$\,erg\,s$^{-1}$, \cep is below the theoretical limit where such a correlation is expected. This behavior is similar to Vela~X-1 and we discuss parallels between the two systems.

\end{abstract}


\section{Introduction}

Neutron star high-mass X-ray binaries (HMXBs), i.e., neutron stars accreting from an early-type stellar companion, typically show very high levels of variability in their X-ray emission.  As the unabsorbed flux is directly related to the mass accretion rate, we can understand these changes as originating from variations of the latter. The physical conditions inside the accretion column, where most of the X-rays are produced, change with accretion rate. By sampling different luminosity levels we can obtain constraints on the geometry and physical processes in the accretion column. In particular, Be-star systems, where the stellar companion has a large circumstellar disk \citep[e.g.][]{okazaki01a}, are ideally suited for studying the luminosity dependence of the X-ray spectrum, as they show weeks- to months-long outbursts which can cover more than two orders of magnitude in luminosity. 

Many accreting neutron stars  show prominent cyclotron resonant scattering features (CRSFs).
CRSFs are produced by resonant scattering of photons off electrons moving   perpendicular to the magnetic field. The electrons are quantized on Landau levels which energies  depend directly on the local magnetic field, and therefore make the observed energy of the CRSF a direct tracer of the B-field \citep[see, e.g.,][and references therein]{schoenherr07a}. With a variable accretion rate the cyclotron line production region can move along the accretion column, sampling different magnetic fields. A precise measurement of the CRSF  therefore probes the accretion geometry of a source.

According to  theoretical calculations,  a strongly asymmetric shape of the fundamental line is predicted, with significant emission wings.
\citep[e.g.][among others]{yahel79a, isenberg98a, araya96a, araya99a, arayagochez00a, schoenherr07a, schwarm10a}. 
The shape is thereby strongly dependent on the accretion geometry and the underlying X-ray continuum.
The asymmetry originates from photon-spawning, where electrons excited to higher harmonic Landau levels cascading back to the ground state and emitting photons
close to the fundamental line energy
  \citep{schoenherr07a}. 

Observationally, however, only very few of the approximately 25 known CRSF sources show any evidence for a deviation from very simply-shaped fundamental lines.
In \gx, \citet{kreykenbohm04a} found marginal evidence for an asymmetric profile at certain pulse phases using \xte data, but the statistics did not allow for a detailed description of the feature. \citet{pottschmidt05a} and \citet{nakajima10a} showed that the fundamental line in \vo is better described by two absorption-like lines at almost the same energy, but with different widths.
\citet{iwakiri12a} claimed the detection of a CRSF in emission rather than absorption at certain phases of  4U~1626$-$67, using \suz data. This feature might lead to a complex line profile in the phase-averaged data, but their data did not provide a high enough quality to measure it.

Among the six CRSF sources studied with the  \textsl{Nuclear Spectroscopic Telescope Array} \citep[\nustar,][]{harrison13a} so far, none show significant deviations from a Gaussian or Lorentzian optical depth profile \citep{herx1, velax1nustar, ks1947, bellm14a, tendulkar14a, bhalerao15a}. A  detailed study of \her revealed good agreement in the line profile between \nustar and \suz, and put the most stringent limits on possible emission wings to date \citep{herx1}

The Be HXMB \cep has a CRSF around 30\,keV, which is ideally suited to be studied in detail with \nustar, thanks to the instruments unprecedented energy resolution as well as increased sensitivity above 10\,keV compared to previous missions. \cep was discovered by \textsl{OSO~7} in 1972 \citep{ulmer72a}  and again detected by \textsl{Ginga} in 1988 \citep{makino88a}. During the 1988 outburst, regular pulsations with a pulse period around 66\,s were discovered, and evidence for a CRSF around 30\,keV was found \citep{koyama91a, mihara91a}. The optical counterpart was identified by \citet{bonnetbidaud98a}, who measured a distance of $3.8\pm0.6$\,kpc.

The most detailed spectral description to date is presented by \citet{mcbride07a}, who used \textsl{Rossi X-ray Timing Explorer} (\xte) data taken during an outburst in 2002. They confirm the CRSF around 30.7\,keV and describe the continuum with an absorbed power-law with a Fermi-Dirac cutoff. By monitoring the source over the outburst they find a hardening of the broadband spectrum with luminosity, but the data quality does not allow the investigation of the dependence of the CRSF energy on luminosity. \citet{mcbride07a} also show that the pulse profile changes significantly as a function of luminosity, confirming the results by \citet{mukerjee00a} who use \xte and {Indian X-ray Astronomy Experiment} ({IXAE}) data.

%
%




The rest of the letter is organized as follows: in Section~\ref{sec:data} we detail the data reduction, and in Section~\ref{sec:spectra} we present the spectral analysis.
Section~\ref{sec:summ} discusses and  summarizes our results  and concludes this letter.

\section{Observations and data reduction}
\label{sec:data}

\begin{figure}
\includegraphics[width=0.95\columnwidth]{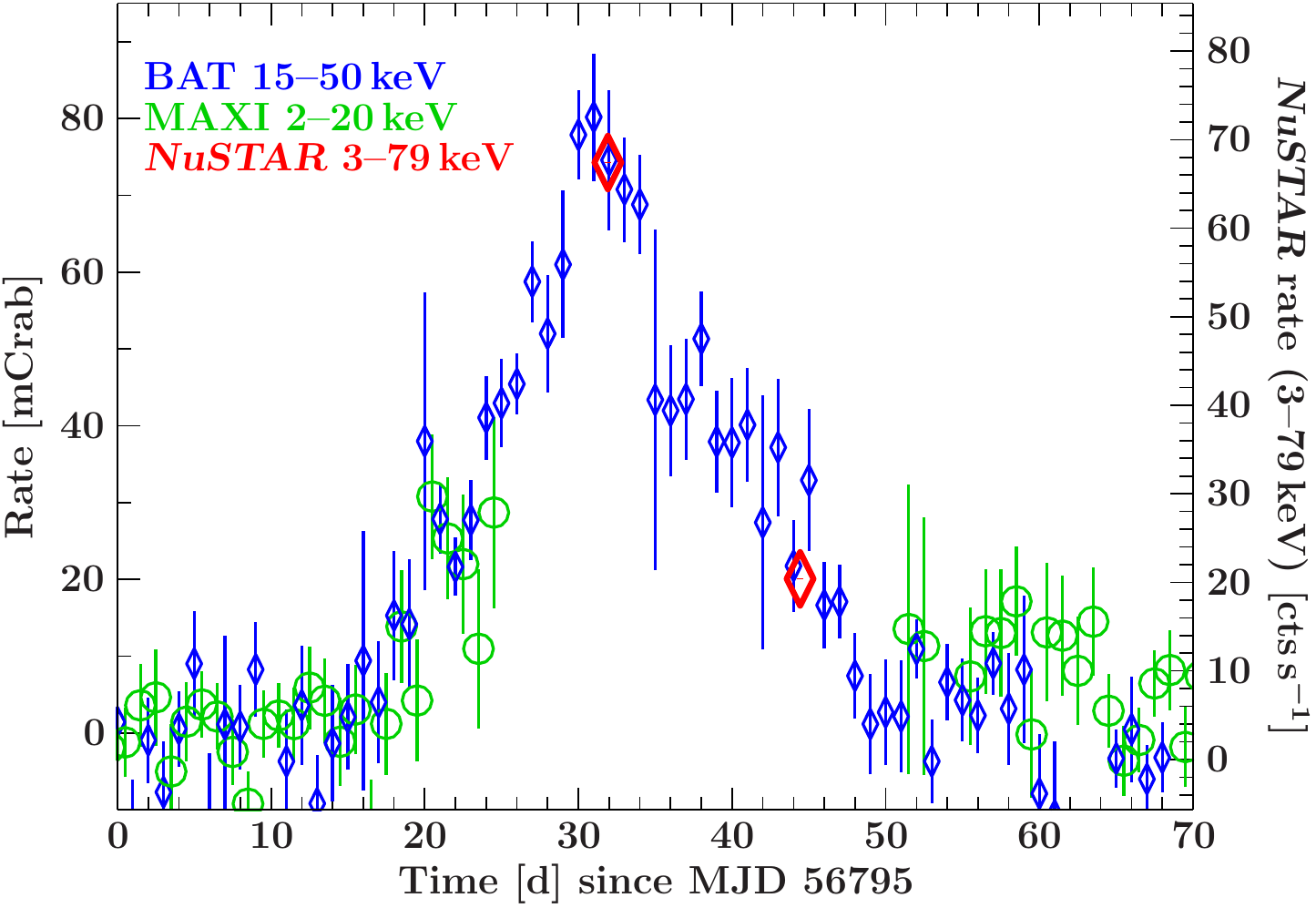}
\caption{Lightcurve of the 2014 outburst of \cep as observed with \swift/BAT (blue diamonds), MAXI (green circles) and \nustar (red diamonds). All rates were rescaled to mCrab in the respective energy band. The right hand $x$-axis gives the measured count-rates for the \nustar data. MAXI did not take data during the maximum of the outburst.}
\label{fig:batlc}
\end{figure}

In June 2014 MAXI detected renewed activity from \cep \citep{nakajima14a}, and \swift performed pointed XRT observations after an automatic BAT trigger \citep{evans14a}. \citet{ozbeyarabaci14a} performed optical observations and found evidence for a strong Be-disk, expected to occur with the onset of the X-ray activity. We triggered \nustar observations and
 observed \cep twice, on 2014 June 18th--19th (MJD 56826.92--56827.84, ObsID 80002016002, observation 1) close to the maximum of the outburst and on 2014 July 1st--2nd  (MJD 56839.43--56840.31, ObsID 80002016004, observation 2) during the decline. Both \nustar observations were supported by \swift/XRT snapshots (ObsIDs 00080436003 and 00080436004, respectively). Figure~\ref{fig:batlc} shows the light curves of \swift/BAT  and MAXI, and the average count-rate of the \nustar observations.

\subsection{\nustar}
We extracted \nustar data  using the standard \texttt{nupipeline} software v1.4.1 as distributed with HEASOFT 6.16. We used CALDB v20150316, taking a time-dependent gain change into account. Standard screening of the data resulted in good exposure times of 40.5\,ks for observation 1 and 41.2\,ks for observation 2. 
We extracted source spectra separately for FPMA and FPMB from a circular region with a radius of $120''$ centered at the J2000 coordinates. The background spectra were extracted from the opposite corner of the \nustar field of view from a circular region with $90''$ radius. 

Spectra were modeled with the Interactive Spectral Interpretation System \citep[ISIS, ][]{houck00a} v1.6.2-30 and errors are given at the 90\% confidence level unless otherwise noted. The data were rebinned within ISIS to a \snr of eight between 3.2--45\,keV and a \snr of three above that, binning at least two channels together.
While the \nustar calibration is nominally good down to 3\,keV we choose to ignore the first few bins in order to allow
 \swift/XRT to drive the model at the lowest energies. 
 We include data up to 60\,keV, where they become background domianted.

\subsection{Swift/XRT}
Data from the \swift X-ray Telescope \citep[XRT,][]{swiftxrtref} were extracted following the steps outlined in the \swift user's guide \citep{swiftxrtguide} using HEASOFT 6.16. Observation 1 (MJD 56826.94--56826.956) was performed in  windowed timing  (WT) mode, and we extracted the source spectrum from a rectangular region with $45''$ length perpendicular to the read-out direction. The background was extracted from similar regions on both sides of the source location. After  standard screening we obtained a good exposure time of 1.0\,ks.

Observation 2 (MJD 56839.462--56839.531) was performed in photon counting (PC) mode and was heavily piled-up. Following the procedure described in the XRT data analysis guide\footnote{\url{http://www.swift.ac.uk/analysis/xrt/pileup.php}} we determined that an annulus extraction region with inner radius   $12''$ and outer radius $60''$  removes most pile-up while still providing a good \snr spectrum.  The background was estimated from a  large region south-east of the source. The observation resulted in 1.6\,ks of good exposure time. 
Both \swift/XRT spectra were rebinned to a \snr of six throughout the used energy range of 0.8--10\,keV.

\section{Spectral analysis}
\label{sec:spectra}

As \citet{mcbride07a} demonstrate, the hard X-ray continuum of \cep is  well described by an absorbed powerlaw with a Fermi-Dirac cutoff \citep{tanaka86a} of the form

\begin{equation}
F(E) \propto E^{-\Gamma} \times \left(1+\exp\left(\frac{E-E_\text{cut}}{E_\text{fold}}\right)\right)^{-1} \quad .
\end{equation}

To that continuum, \citet{mcbride07a} add a CRSF modeled by a multiplicative absorption line\footnote{We note that while a CRSF is produced by resonant scattering, not
absorption, a possible parametrization is the same as for the latter.}  with a Gaussian optical depth profile (\texttt{gabs} in XSPEC) and a narrow additive fluorescent \feka line. We apply this same model to both observations separately, modeling \swift/XRT, \nustar/FPMA, and FPMB simultaneously  and allowing for a cross-calibration constant relative to FPMA for FPMB and XRT ($\text{C}_\text{FPMB}$ and $\text{C}_\text{XRT}$). We use the \texttt{phabs} absorption model with abundances by \citet{wilms00a} and cross-sections by \citet{verner96a}. 

\begin{figure}
\includegraphics[width=0.95\columnwidth]{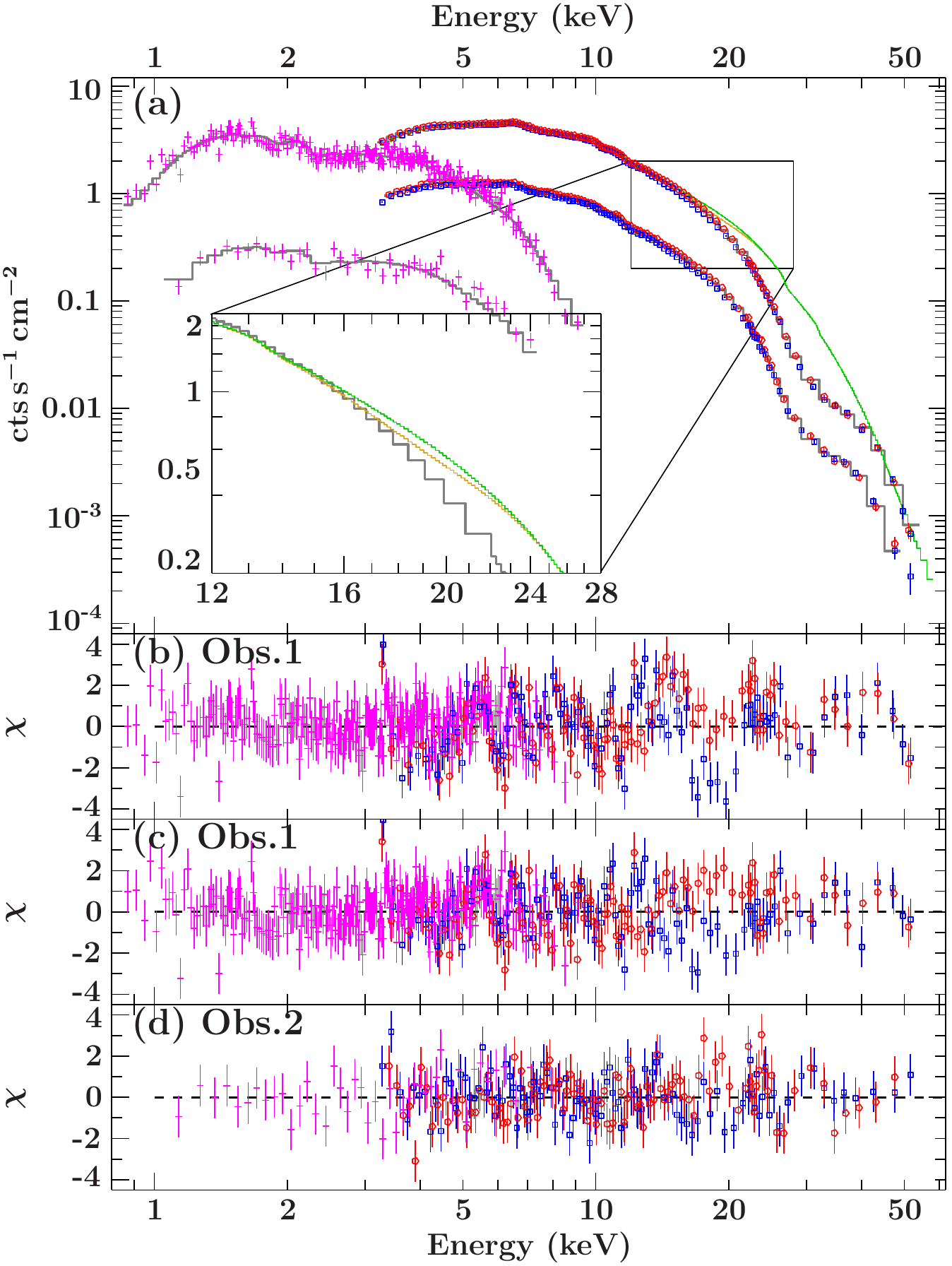}
\caption{\textit{(a)} Count spectra and best-fit models of both observations of \cep. \swift/XRT data are shown in magenta, \nustar/FPMA data in red, and \nustar/FPMB data in blue. The best-fit model is shown in gray, the model evaluated without the absorption lines in green, and the model with only the primary CRSF in orange. The inset shows a zoom on the red edge of the CRSF in observation 1. \textit{(b)} residuals of observation 1 to a FDcut model with only one absorption line, \textit{(c)} residuals of observation 1 to the best-fit model, \textit{(d)} residuals of observation 2 to the best-fit model. Data were rebinned for plotting purposes.}
\label{fig:spec}
\end{figure}

The model results in an unacceptable fit for both observations ( $\chi^2/\text{dof} = 2275/1087 =2.09$ for observation 1 and $\chi^2/\text{dof} = 1128/675 = 1.67$ for observation 2)
 with strong residuals below 10\,keV. We therefore add a blackbody component with $kT_\text{BB}\approx 1$\,keV  which improves the fit significantly. We obtain $\chi^2/\text{dof} =1324/1085 =1.22$ for observation 1 and $\chi^2/\text{dof} = 802/673 = 1.19$ for observation 2. The residuals of observation 1 for this model are shown in Figure~\ref{fig:spec}\textit{b}. 
 
 As can be seen in Figure~\ref{fig:spec}\textit{b}, the residuals still show some structure between 10--20\,keV which is not modeled by the CRSF. We therefore add another multiplicative absorption  line with a Gaussian optical depth profile. This addition provides a significant improvement  and results in a good fit, with $\chi^2/\text{dof} = 1215/1082 =1.12$ for observation 1 and $\chi^2/\text{dof} = 776/670 =1.16$ for observation 2. The energy of the CRSF does not change significantly when adding the second absorption model, which is found to be around $E_\text{abs} \approx 19$\,keV in both observations. The residuals to this best-fit model are shown in Figure~\ref{fig:spec}\textit{c} and \textit{d} for observation 1 and 2 respectively, and the best-fit parameters are given Table~\ref{tab:spec}.

\begin{deluxetable}{rll}
\tablewidth{0pc}
\tablecaption{Parameters of the best-fit Fermi-Dirac cutoff model for both observations.\label{tab:spec}}
\tablehead{\colhead{Parameter} & \colhead{Obs. 1} & \colhead{Obs. 2} }
\startdata
 $ N_\text{H}~[10^{22}\,\text{cm}^{-2}]$ & $1.05^{+0.11}_{-0.12}$ & $1.41\pm0.25$ \\
 $ A_\text{cont}\tablenotemark{a}$ & $0.061^{+0.008}_{-0.010}$ & $0.021^{+0.004}_{-0.005}$ \\
 $ \Gamma$ & $0.83^{+0.07}_{-0.11}$ & $0.96^{+0.09}_{-0.14}$ \\
 $ E_\text{cut}~(\text{keV})$ & $24\pm4$ & $25\pm4$ \\
 $ E_\text{fold}~(\text{keV})$ & $5.7^{+0.5}_{-0.6}$ & $5.7^{+0.6}_{-0.8}$ \\
 $ E_\text{CRSF}~(\text{keV})$ & $30.39^{+0.17}_{-0.14}$ & $29.42^{+0.27}_{-0.24}$ \\
 $ \sigma_\text{CRSF}~(\text{keV})$ & $5.8\pm0.4$ & $4.9\pm0.4$ \\
 $ d_\text{CSRF}\tablenotemark{b}~(\text{keV})$ & $20^{+5}_{-4}$ & $16.6^{+4.0}_{-3.0}$ \\
 $ E_\text{abs}~(\text{keV})$ & $19.0^{+0.5}_{-0.4}$ & $18.5\pm0.7$ \\
 $ \sigma_\text{abs}~(\text{keV})$ & $2.5\pm0.4$ & $2.1\pm0.5$ \\
 $ d_\text{abs}\tablenotemark{b}~(\text{keV})$ & $0.60^{+0.24}_{-0.17}$ & $0.37^{+0.21}_{-0.15}$ \\
 $ A(\text{Fe\,K}\alpha)\tablenotemark{a}$ & $\left(1.39^{+0.16}_{-0.14}\right)\times10^{-3}$ & $\left(2.8^{+0.8}_{-0.6}\right)\times10^{-4}$ \\
 $ \sigma(\text{Fe\,K}\alpha)~(\text{keV})$ & $0.42\pm0.05$ & $0.34^{+0.12}_{-0.10}$ \\
 $ E(\text{Fe\,K}\alpha) (\text{keV})$ & $6.474^{+0.030}_{-0.032}$ & $6.39^{+0.06}_{-0.07}$ \\
 $ A_\text{BB}\tablenotemark{c}$ & $\left(2.22^{+0.41}_{-0.29}\right)\times10^{-3}$ & $\left(7.3^{+1.7}_{-1.3}\right)\times10^{-4}$ \\
 $ kT_\text{BB}~(\text{keV})$ & $0.899^{+0.030}_{-0.031}$ & $0.96\pm0.06$ \\
 $ \text{C}_\text{FPMB}$ & $1.0319\pm0.0019$ & $1.023\pm0.004$ \\
 $ \text{C}_\text{XRT}$ & $0.962\pm0.019$ & $0.91\pm0.05$ \\
\hline$\chi^2/\text{d.o.f.}$   & 1215.83/1082& 776.35/670\\$\chi^2_\text{red}$   & 1.124& 1.159\enddata
\tablenotetext{a}{in photons\,keV$^{-1}$\,s$^{-1}$\,cm$^{-2}$ at 1\,keV}\tablenotetext{b}{line depth, optical depth = $d/(\sigma\sqrt{2\pi})$}\tablenotetext{c}{in $10^{39}$\,erg\,s$^{-1}$ for a source at 10\,kpc}
\end{deluxetable}

Using two lines with Gaussian optical depth profiles describes the shape of the CRSF in both observations very well. Using a pseudo-Lorentzian  profile instead (modeled by the \texttt{cyclabs} model) results in a very similar fit and also requires a second absorption feature. The line energies are about 2\,keV lower, consistent with the expected difference between the models \citep[see, e.g.][]{staubert14a}. Our model is therefore insensitive to the slight differences in shape between the two profiles, similar to results obtained for Her~X-1 \citep{herx1} and V\,0332+35 \citep{pottschmidt05a, nakajima10a}.

The CRSF energy shows small but statistically significant variations with luminosity, decreasing from $E_\text{Obs 1}=30.39^{+0.17}_{-0.14}$\,keV in the first observation to $E_\text{Obs 2} = 29.42^{+0.27}_{-0.24}$\,keV in the second observation. This behavior is qualitatively the same with or without the second absorption feature.
In order to rule out the possibility that small changes in the continuum parameters influence the measured energy of the CRSF, we performed a simultaneous fit of both datasets. This was possible as the continuum parameters (besides the normalization) do not change significantly between the two observations. In the simultaneous fit,  the photon index $\Gamma$, the folding energy $E_\text{fold}$, the cutoff energy $E_\text{cut}$, the absorption column $\nh$, and the \feka line parameters are tied between both datasets. We obtain very similar results with respect to the two absorption features and the black body component, in particular the luminosity dependence of the CRSF energy is seen with the same significance and both observations require a second absorption feature.

In order to further investigate the choice of the continuum on the shape and  energy of the CRSF, we also model the data with an NPEX model \citep{mihara95b}, in which we fix the secondary power-law index to $\Gamma_2= -2$. We obtain a very similar statistical quality of fit and consistent parameters. The energy of the line is fitted to $E_\text{Obs 1}=30.59^{+0.13}_{-0.15}$\,keV and $E_\text{Obs 2}=29.43^{+0.23}_{-0.22}$\,keV. The secondary absorption feature is also clearly visible in the residuals and its addition leads to a similar improvement in terms of $\chi^2$. 

\section{Discussion}
\label{sec:summ}
We have presented simultaneous \swift and \nustar observations at two different luminosities taken during an outburst of \cep in June and July 2014. We describe the broad-band spectrum with a power-law attenuated by absorption and a Fermi-Dirac cutoff. We find that the continuum does not change significantly between the observations but that the CRSF shows complex behavior.

\subsection{Shape of the CRSF}
We have shown that to accurately describe the hard X-ray spectrum of \cep we require a second absorption feature in addition to the prominent CRSF around 30\,keV. If this feature were  the fundamental line, it would imply a lower magnetic field strength than previously suggested. That is, however, unlikely as we expect the ratio of the fundamental and first harmonic line energy to be close to 2, while we measure $\approx 1.56$.  Other sources also show deviations from the expected factor, however, they are typically much smaller \citep[$<10\%$, e.g.,][]{pottschmidt05a, mueller13a}. The deviation we observe is too large to be explained by relativistic effects \citep{meszaros92a}.

The second feature is also unlikely to result from sampling
different accretion columns or different regions.
If that were the case, the strength of the 30\,keV line would be difficult to explain as this scenario implies that  the line should only be present at some phase intervals. 
In order to further investigate this we will present results of phase-resolved spectroscopy in a separate paper (Bhalerao et al., in prep.). The production of two CRSFs of such different energies in different accretion columns would also indicate a strong deviation from a simple dipole magnetic field, which is not expected \citep[see][for a discussion of the influence of multipole fields on the CRSF]{nishimura05a}.

 \begin{figure}
\includegraphics[width=0.95\columnwidth]{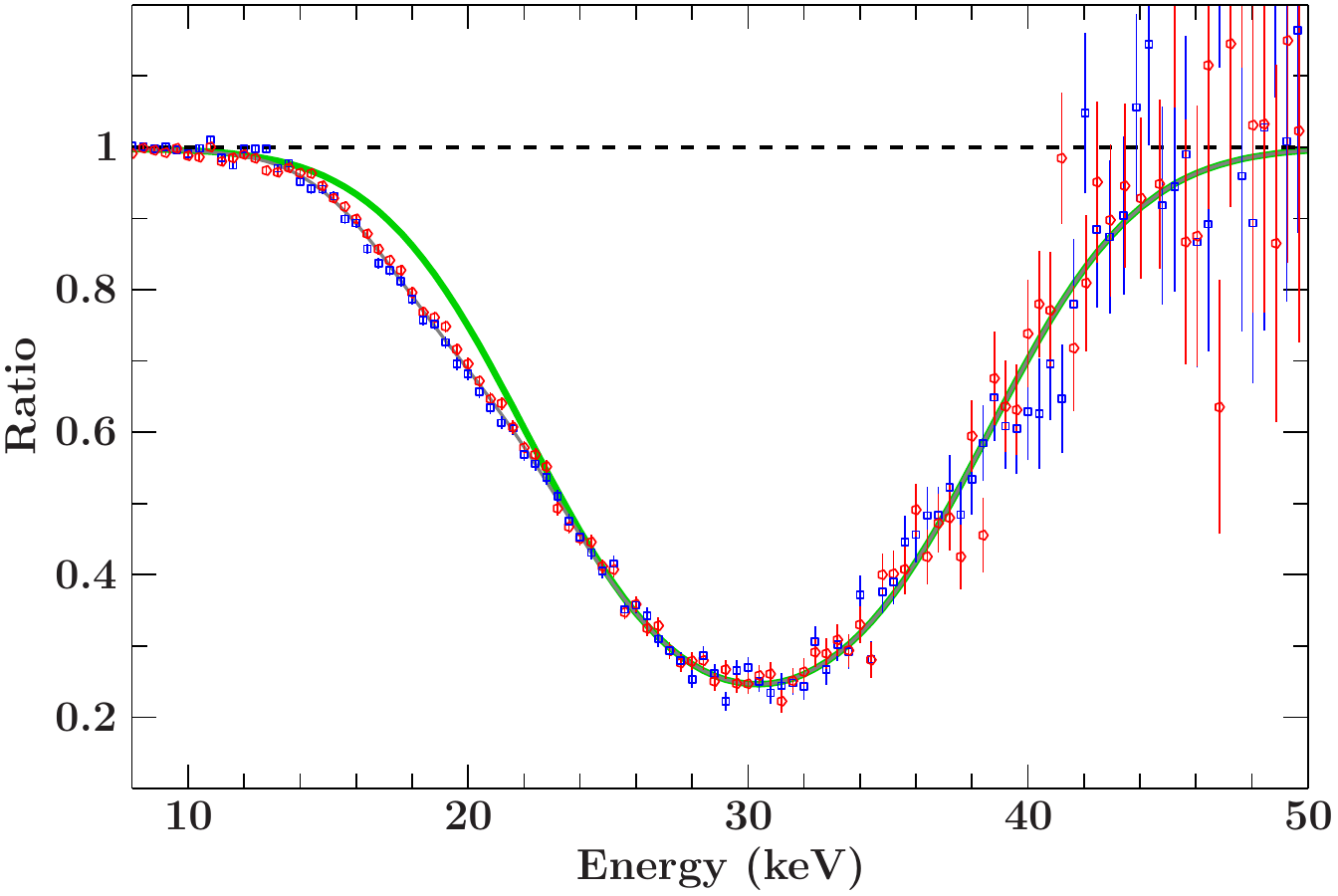}
\caption{Residuals of observation 1 in terms of data-to-model ratio. The ratio was calculated by setting the strength of both absorption features to zero. FPMA data are shown in red, FPMB in blue. The best-fit model is shown in gray. The green line shows the ratio when only setting the second absorption feature to zero, i.e., when assuming a symmetric line profile. The data were strongly rebinned for the plot. }
\label{fig:lineshape}
\end{figure}

The most likely explanation for the second absorption feature is a deviation of the shape of the CRSF from a smooth Gaussian or Lorentzian profile.  
This deviation is highlighted in Figure~\ref{fig:lineshape}, where we plot the data-to-model ratio using the best-fit model with both lines removed. We superimpose the ratio between this model and the best-fit model with the secondary feature removed, which implicitly shows a simple symmetric line profile. From this ratio the data clearly deviate at the red wing.
A similar deviation might be present on the blue side, however, the data quality does not us allow to constrain this. Adding a feature there with a similar optical depth does not change the statistical quality of the fit significantly.

Prominent emission wings would be the most obvious explanation for deviations from a smooth line. However, such emission wings typically require a harder spectrum than we observe to spawn enough photons to become significant \citep{schoenherr07a}. On the other hand, depending on geometry and optical depth, photon spawning can lead to distorted shapes of the scattering trough without creating measurable emission wings \citep{schwarm10a}.
As can be seen in the inset of Figure~\ref{fig:spec} and in Figure~\ref{fig:lineshape}, the second absorption feature is located exactly at the energy where the primary CRSF starts to produce significant deviations from the continuum spectrum, supporting this interpretation.


\subsection{Luminosity dependence of the CRSF}
\begin{figure}
\includegraphics[width=0.95\columnwidth]{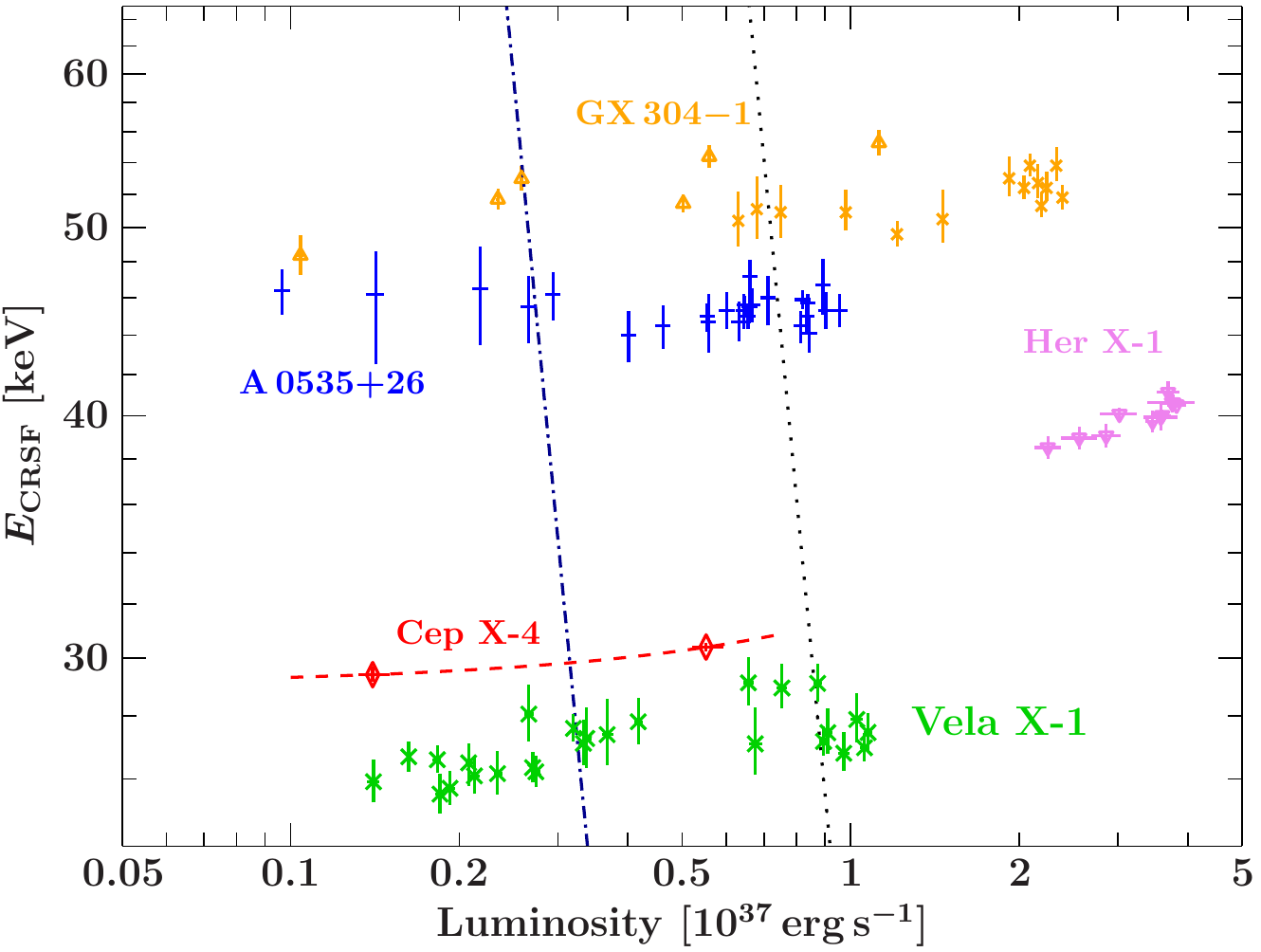}
\caption{Dependence of the CRSF energy on luminosity for different sources. \cep data are plotted as red diamonds  with a best-fit linear correlation superimposed (dashed line). \vela data (green asterisk) are from \citet{velax1nustar} and show the energy of the first harmonic CRSF divided by 2. A\,0535+26 data (blue crosses) are from \citet{caballero07a}, Her~X-1 data (magenta down-pointing triangles) are from \citet{staubert07a}, and GX~304$-$1 (orange up pointing triangles) are from \citet{yamamoto11a} and \citet{klochkov12a}. The dotted line indicates the original Coulomb stopping luminosity presented by \citet{becker12a} and the dot-dashed line indicates the updated values for \vela \citep{velax1nustar}.}
\label{fig:ecrsf2lx}
\end{figure}

Thanks to \nustar's energy resolution, we are able to constrain the centroid of the line energy to better than $\pm0.3$\,keV. This allowed us to measure a significant change between the two observations, which appears to be correlated with luminosity. 
To highlight that correlation and put it into context, we show the CRSF energy as a function of luminosity in Figure~\ref{fig:ecrsf2lx} for different sources.
The energy decreases with declining flux, leading to a positive correlation between luminosity and CRSF energy. The luminosity of \cep is  between (1--6)$\times10^{36}$\,erg\,s$^{-1}$ for a distance of 3.8\,kpc, which is below the theoretical limit for the formation of a shock  in the accretion column \citep{becker12a}. We  therefore expect a constant CRSF energy uncorrelated with luminosity, as observed in A\,0535+26 \citep[but see \citealt{sartore15a} for indications of a correlation]{caballero07a}. 
In recent \nustar observations of \vela, \citet{velax1nustar}  show that this source also has a significant positive correlation between CRSF energy and luminosity at luminosities below $10^{37}$\,erg\,s$^{-1}$. 

\citet{velax1nustar}  follow the calculations of \citet{becker12a} but  allow for wind-accretion instead of disk-accretion and assume a massive neutron star (around 2\,\msun in \vela). This results in a narrower accretion column than assumed in \citet{becker12a}, which in turn decreases the luminosity threshold required for shock formation significantly, moving the measured luminosities of \vela partially above it (see the dot-dashed line in Figure~\ref{fig:ecrsf2lx}). \cep also crosses this adopted line, although the mass of the neutron star in this system is not known. Additionally,  \cep is a Be-system, and so we expect that the accreted matter forms a temporary accretion disk around the neutron star, which leads to a different accretion geometry than in purely wind accreting systems \citep[e.g.][]{ghosh79a, okazaki13a}. If the accretion geometry could be constrained, this correlation might indicate that the neutron star in \cep is also massive, with $M>2\,\msun$.

\subsection{Summary}
\nustar has revealed two new interesting features of the CRSF in \cep: a distorted profile and a luminosity dependence of the line's energy. This makes \cep the first system where a significant deviation from a symmetric line profile has been measured in the phase averaged spectrum and the second system where a positive correlation between CRSF energy and luminosity has been found at luminosities below $10^{37}$\,erg\,s$^{-1}$. The latter discovery challenges the current understanding of the accretion column  geometry, as the line forming region is expected to be at the neutron star surface and therefore independent of luminosity. 

The discovery of a complex line profile on the other hand is a good qualitative confirmation of theoretically predicted line-profiles when taking photon-spawning and magnetic field gradients into account \citep[e.g.][]{nishimura05a,schoenherr07a}. By combining detailed calculations of the line profile with sophisticated light bending calculations, it should be possible to improve our understanding of the emission geometry in \cep and other neutron star systems.

\acknowledgments
We thank the anonymous referee for the valuable comments.
This work was supported under NASA Contract No. NNG08FD60C, and
made use of data from the {\it NuSTAR} mission, a project led by
the California Institute of Technology, managed by the Jet Propulsion
Laboratory, and funded by the National Aeronautics and Space
Administration. We thank the {\it NuSTAR} Operations, Software and
Calibration teams for support with the execution and analysis of these observations. This research has made use of the {\it NuSTAR}
Data Analysis Software (NuSTARDAS) jointly developed by the ASI
Science Data Center (ASDC, Italy) and the California Institute of
Technology (USA). Support for this work was provided by NASA through the Smithsonian Astrophysical Observatory (SAO) contract SV3-73016 to MIT for Support of the Chandra X-Ray Center (CXC) and Science Instruments; CXC is operated by SAO for and on behalf of NASA under contract NAS8-03060.


%

\end{document}